\documentclass[twocolumn,tightenlines,superscriptaddress,showpacs,amsmath,amssymb,aps,prb,floatfix]{revtex4}

\usepackage{graphicx}
\usepackage{dcolumn}
\usepackage{bm}

\usepackage[usenames,dvipsnames]{xcolor}
\usepackage{dsfont}
\usepackage{amsbsy}

\newcommand{\Eq}[1]{Eq.~(\ref{#1})}
\newcommand{\Fig}[1]{Fig.~\ref{#1}}

\newcommand{\App}[1]{Appendix~\ref{#1}}

\newcommand{\ket}[1]{\left|{#1}\right\rangle}

\newcommand{\bra}[1]{\left\langle{#1}\right|}

\newcommand{\tsc}[1]{\textsuperscript{#1}}

\begin{document}

\title{Keeping a spin qubit alive in natural silicon:\\ Comparing optimal working points and dynamical decoupling}

\author{S.~J.~Balian}
\email{s.balian@ucl.ac.uk}
\affiliation{Department of Physics and Astronomy, University College London,
Gower Street, London WC1E 6BT, United Kingdom}

\author{Ren-Bao~Liu}
\affiliation{Department of Physics, The Chinese University of Hong Kong, Hong Kong, China}

\author{T.~S.~Monteiro}
\affiliation{Department of Physics and Astronomy, University College London,
Gower Street, London WC1E 6BT, United Kingdom}

\date{\today}

\begin{abstract}

There are two distinct techniques of proven effectiveness for extending the
coherence lifetime of spin qubits in environments of other spins.
One is dynamical decoupling, whereby the qubit is subjected to a 
carefully timed sequence of control pulses;
the other is tuning the qubit towards `optimal working points' (OWPs),
which are sweet-spots for reduced decoherence in magnetic fields.
By means of quantum many-body calculations,
we investigate the effects of dynamical decoupling pulse sequences far from and near OWPs
for a central donor qubit subject to decoherence from a nuclear spin bath.
Key to understanding the behavior is to analyse the degree of suppression of the
usually dominant contribution from independent pairs of flip-flopping spins
within the many-body quantum bath. We find that to simulate recently measured Hahn echo decays at OWPs (lowest-order dynamical decoupling),
one must consider clusters of three interacting spins, since independent pairs do not even give finite $T_2$ decay times.
We show that while operating near OWPs, dynamical decoupling sequences require
hundreds of pulses for a single order of magnitude enhancement of $T_2$,
in contrast to regimes far from OWPs, where only about ten pulses
are required.

\end{abstract}


\maketitle

\section{Introduction}

Individual electronic and nuclear spins are promising candidates for realizing scalable
quantum computing in solid state systems such as silicon.\cite{Zwanenburg2013} 
In addition, they offer a valuable test-bed for the experimental investigation
of decoherence driven by quantum baths which typically comprise large numbers of
spins surrounding a central spin qubit.
Despite the large number of bath spins involved,
accurate simulation of experimental coherence decays is computationally tractable since the bath can
often be decomposed into independent contributions from many small sets or clusters of spins.\cite{DeSousa2003_1,Witzel2005,Witzel2006,Yao2006}
In many cases, the dominant contribution to the decoherence dynamics arises from {\em pairs} of bath spins; in effect, from the magnetic noise due to the independent
`flip-flopping' of spin pairs. Contributions from larger clusters are usually only needed for high accuracy;\cite{Witzel2010,Witzel2012,Zhao2012}
however, there is also interest in cases where experimental evidence for the many-body or large-cluster nature of
the bath might be most evident. For example, it was recently found that applying dynamical decoupling sequences of microwave pulses
may enhance many-body correlations, relative to the independent spin-pair contribution.\cite{Ma2014}

A defining characteristic of a quantum bath is the back-action between the central system and environment,
making the bath dynamics sensitive to the state of the central spin.
A particularly striking example has been identified recently in spin systems with special coherence sweet-spots
termed `optimal working points' (OWPs),\cite{Mohammady2010,Mohammady2012,Balian2012,Wolfowicz2013,Balian2014}
where theory and experiment found
that coherence times can change by orders of magnitude with even small variations
in applied magnetic field ($\sim 200$ G).
A drastic change in back-action occurs via changes to the states of the central spin only:
the change in external field has little direct effect on the bath spin-pair dynamics.
However, it was found that usual cluster simulations using only pairs predicted infinite coherence times for the lowest order dynamical decoupling
sequence (Hahn spin echo), at odds with experiment.\cite{Balian2014}

These findings motivate us to investigate here the interplay between dynamical decoupling and OWPs
for a quantum bath system and in particular, to clarify where and to what extent, the independent spin-pair contribution
dominates. Dynamical decoupling is one of the most established methods for extending coherence.\cite{Carr1954,Meiboom1958,Viola1998,Viola1999,Uhrig2007,Witzel2007,Witzel2007_2,Witzel2008,Yang2008_1,Biercuk2009,Preskill2011}
It involves subjecting the qubit spin
to a sequence of microwave or radio pulses.
A wide variety of solid state spin qubits have been studied under dynamical decoupling control; these include
Group V donors in silicon,\cite{Tyryshkin2006,Tyryshkin2010,Wang2011,Pla2012,Pla2013,Wang2012,Steger2012,Saeedi2013,Ma2014,Muhonen2014}
nitrogen vacancy centres in diamond,\cite{DeLange2010,Zhao2012,Pham2012,Wang2012_1}
GaAs quantum dots,\cite{Zhang2008}
rare-earth dopants in silicates,\cite{Fraval2005} malonic acid crystals\cite{Du2009}
and adamantane.\cite{Peng2011}

Long-lived coherence is a key requirement
for implementing fault-tolerant quantum computation,\cite{Shor1996}
as well as quantum memory.\cite{Biercuk2009}
One way of extending coherence times $T_2$ in silicon is to use
isotopically enriched samples in which the abundance
of nuclear spin isotopes is significantly reduced.\cite{Tyryshkin2012,Wolfowicz2012,Saeedi2013}
However, it is advantageous to retain nuclear spins for their potential use as
long-lived quantum registers.\cite{Pla2014}
It is thus also of practical importance to understand whether 
dynamical decoupling and OWP techniques may be advantageously combined for a quantum bath
of nuclear spins.
For donor electronic qubits in silicon, it is known that due to inhomogeneous broadening from naturally-occurring $^{29}$Si spin isotopes,
there is a significant gap between the $T_2 \sim 100$~ms
in natural silicon near an OWP,\cite{Wolfowicz2013,Balian2014} and the $T_2\sim 2$~s in isotopically enriched $^{28}$Si with a low donor concentration
at the same OWP.\cite{Wolfowicz2013} Also, dynamical decoupling may be useful when it is convenient to operate with the magnetic field close to but not exactly at the OWP.

It is well established that for dynamical decoupling to be effective, the pulse spacing
$\tau=t/2N$ for a sequence of $N$ control pulses (where $t$ is the total evolution time)
cannot exceed the correlation time of the bath noise.
But the relevant correlation time, in turn, is an emergent property of the underlying
microscopic quantum bath, comprised of typically $\sim 10^4-10^5$ significant clusters of spins
of different coupling strengths, different sizes and subject to
varying degrees of back-action from the central qubit. Therefore,
to quantitatively simulate the response to dynamical decoupling, a realistic simulation
of the combined system-bath dynamics at the microscopic level is important. 

\begin{figure*}[t]
\includegraphics[width=7.0in]{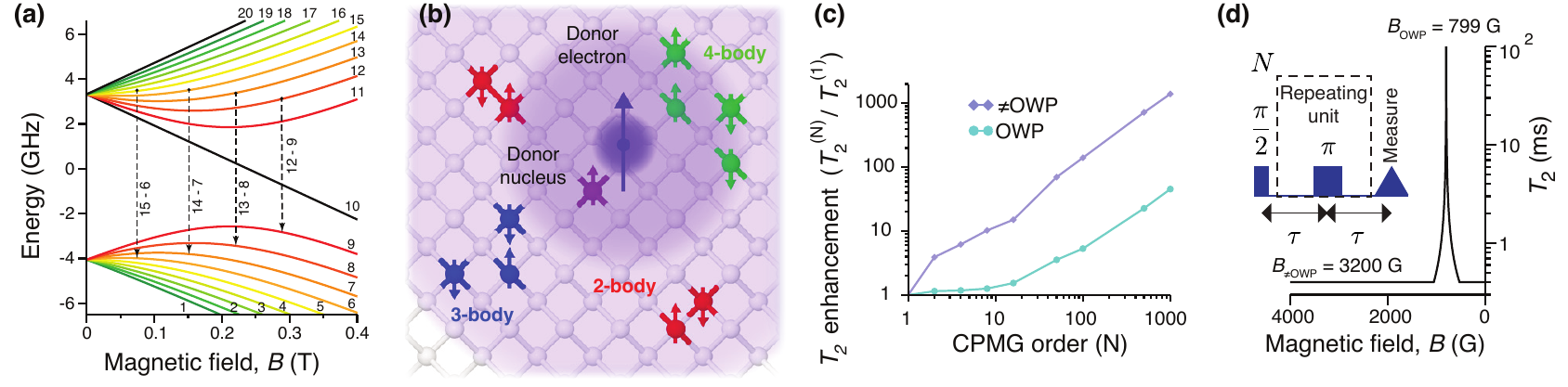}
\caption{(color online)
{\bf(a)} The spectra of donor spin systems such as arsenic, antimony or bismuth (pictured) 
are affected by strong mixing between the electron and host nuclear spin, at magnetic fields $B$
smaller or comparable to the hyperfine coupling $A$, allowing a richer behavior than 
unmixed electron spins. At particular field values termed optimal working points (OWPs),
decoherence can be strongly suppressed; the arrows indicate the transitions with four of the most significant OWPs.
{\bf(b)} Coherences of the central electronic spins are dephased primarily by a surrounding 
quantum bath of clusters of 2, 3, 4 or more nuclear spin impurities (for natural silicon, pictured) 
or other donors (for isotopically enriched silicon).
{\bf(c)} Effect of dynamical decoupling (CPMG with even pulse numbers $N$) as $N \to \infty$.
Plots $T^{(N)}_2/T_2^{(1)}$ showing enhancement of the electron spin coherence time $T_2$ as a function of pulse number $N$,
relative to the $N=1$ Hahn echo value. We find that while dynamical decoupling far from the OWP enhances $T_2$ by an order of
magnitude with about 10 pulses, in contrast, behavior close to an OWP is insensitive to dynamical decoupling for low $N$.
For high $N$, the behaviors near and far from OWPs become comparable.
{\bf(d)} Illustrates coherence enhancement as $B \to B_{\textrm {OWP}}$ (the Hahn spin echo time $T^{(1)}_2$ is plotted). 
The OWP is for a bismuth donor in natural silicon, investigated 
experimentally in Refs.~\onlinecite{Wolfowicz2013,Balian2014}. {\bf Inset of (d):} The CPMG dynamical decoupling sequence consists of
the initial $\pi/2$ pulse, followed by the $-\tau-\pi-\tau-$echo sequence
repeated $N$ times. The coherence times in (c) are when the CPMG decays in \Fig{Fig:Decays} and
the fits to the decays in \Fig{Fig:HighCPMG} have fallen to $1/e$.
OWP results are for the $\ket{14} \to \ket{7}$ transition for which $B_{\textrm {OWP}} = 799$~G.
For the field value near the OWP ($B=795$~G) in (c), $T_2^{(1)} \simeq 96$~ms
while $T_2^{(1)}\simeq 0.79$~ms in the $\ne$OWP regime ($B=3200$~G).
The OWP curve in (d) was calculated using the analytical formula \Eq{Eq:T2eq}.}
\label{Fig:Entry}
\end{figure*}

In this paper, we present quantum many-body simulations of the system-bath dynamics using the cluster correlation expansion (CCE),\cite{Yao2006,Yang2008_2008E_2009}
including contributions from clusters of up to 5 spins (CCE5) as illustrated in \Fig{Fig:Entry}(b).
We compare coherence decays at an OWP with regimes far from an OWP (denoted by `$\ne$OWP')
with a view to identify the optimal strategy for enhancing electronic spin coherence times of donor spins
in natural silicon. Simple analytical expressions for the behavior of independent bath pairs coupled to the qubit
aid understanding in all the regimes we consider.

In addition, our work fills an outstanding gap in understanding decoherence near OWPs.
It was recently reported that near OWPs, numerical calculations of the Hahn spin echo ($N=1$) involving only independent pair dynamics (CCE2)
yield results in conflict with experiments: coherences decay initially, then after a short time, the decays stop.\cite{Balian2014}
In the present work, we undertake the computationally more challenging CCE$3-5$ many-body calculations 
in order to clarify the origin of the measured coherence decays.
We find that three-spin clusters suffice to give decays in good agreement with previously reported experimental results.\cite{Wolfowicz2013}

It is worth clarifying the physical meaning of the above-mentioned three-cluster result. It is not a matter of enlarging the
quantum bath with additional nuclear spin clusters of the same size. As illustrated in \Fig{Fig:Entry}(b),
a three-spin cluster (blue) can be decomposed into three distinct flip-flopping pairs (each nuclear spin 
can contribute to more than one flip-flopping pair). Put simply, if all such three-clusters 
in a given, randomly generated set of impurities in a crystal are decomposed into the constituent flip-flopping pairs,
an infinite decay time is obtained. If, however, the exact same configuration of spin impurities are aggregated into the `triangle' structures
illustrated in \Fig{Fig:Entry}(b),
the correct experimental behavior emerges.
To our knowledge, there is no other example of a central spin system which so
fully eliminates the pair-driven dynamics.

In contrast, for $\ne$OWP regimes, Hahn decays for dipole-allowed transitions are well described by CCE2.\cite{Witzel2006,Ma2014}
However, for modest $N \lesssim 10$ pulse numbers there can be a large correction to CCE2
from clusters of $3-4$ spins and for even $N$, but CCE2 still
gives $T_2$ on the correct experimental timescale.\cite{Ma2014}
For larger $N$, we find that CCE numerics including only independent pairs (CCE2) once again gives converged decays in all regimes whether
in OWP or $\ne$OWP regimes, so many-body calculations become progressively less important as $N \to \infty$.

The paper is organized as follows. In Section~II we review the theory of decoherence and mixed donor qubits
possessing OWPs. Our main results are presented in Section~III,
first for the Hahn spin echo, then for dynamical decoupling with moderate $N$.
In Section~IV, we provide an analysis on the suppression of the independent pairs contribution to decoherence
as $N \to \infty$, as well as the magnetic field approaching the OWP: $B \to B_{\textrm{OWP}}$.
Finally, we present large $N$ results in Section~V and discuss important aspects of our results in Section~VI.

For our dynamical decoupling calculations, we have chosen the Carr-Purcell-Meiboom-Gill (CPMG) sequence which
applies a set of $N$ periodically spaced near-instantaneous pulses (CPMG$N$) as illustrated
in \Fig{Fig:Entry}(d).\cite{Carr1954,Meiboom1958,Witzel2007}

\section{Decoherence as qubit-bath entanglement}

\subsection{Overlap of conditional bath evolutions}

For quantum baths, decoherence of a central spin system is understood in terms of entanglement 
between the system and the bath, as the combined system evolves under a total
Hamiltonian given by:
\begin{equation}
 \hat{H}_{\text{tot}} =
 \hat{H}_{\text{CS}}
 + \hat{H}_{\text{int}}
 + \hat{H}_{\text{bath}},
\label{Eq:Htotal}
\end{equation}
where $\hat{H}_{\text{CS}}$ is the central system Hamiltonian
including all internal nuclear and electronic degrees of freedom,
while $\hat{H}_{\text{bath}}$ is the bath Hamiltonian and $\hat{H}_{\text{int}}$
describes the system-bath interaction.

We start by placing the initial system state in a coherent superposition
of an upper state $\ket{i=u}$ and a lower state $\ket{i=l}$ by applying a $\pi/2$ pulse:
$\ket{\Psi(t=0)} = \tfrac{1}{\sqrt{2}} \left( \ket{u}+\ket{l} \right) \otimes \ket{\mathcal{B}(t=0)}$,
where $\ket{\mathcal{B}(t=0)}$ is a product state of the eigenstates of the non-interacting bath.
Under the joint system-bath dynamics, the initial state evolves into an entangled state:
\begin{equation}
\sqrt{2}\ket{\Psi(t)} = \left( e^{-iE_ut} \ket{u} \otimes \ket{\mathcal{B}_u(t)} + e^{-iE_lt} \ket{l} \otimes \ket{\mathcal{B}_l(t)} \right),
\label{Eq:tangle}
\end{equation}
where $\ket{\mathcal{B}_{u,l}(t)} = \hat{T}_{u,l} \ket{\mathcal{B}(t=0)}$
with unitaries $\hat{T}_{u,l}$ for each of the two system levels, and we have assumed no depolarization of
the central states during the evolution.
The measured temporal coherence decays $|\mathcal{L}(t)|$ can be simulated if one can 
accurately calculate the resultant overlap between the bath states correlated with the upper and lower
qubit states:
\begin{equation}
|\mathcal{L}(t)  |\propto |\langle\mathcal{B}_u(t)|\mathcal{B}_l(t)\rangle| =
|\langle\mathcal{B}(0) | \hat{T}_u^\dagger \hat{T}_l | \mathcal{B}(0)\rangle|.
\label{Eq:unitary1}
\end{equation}
Since the initial bath states are usual trivial thermal spin states, the challenge is
to obtain the corresponding unitaries $\hat{T}_{u,l}$
for extremely large baths ($\approx 10^4$ spins for convergent CCE).

OWPs are sweet-spots in $B$-field values where the two unitaries involving the upper and lower levels equalise:
$\hat{T}_l  \simeq \hat{T}_u$, occurring when $P_u \simeq P_l$.\cite{Balian2014}
This means that the state given by \Eq{Eq:tangle} can be written as
\begin{equation}
\sqrt{2}\ket{\Psi(t)}=
\left( e^{-iE_ut} \ket{u} + e^{-iE_lt} \ket{l}\right) \otimes \hat{T}_u(t) \ket{\mathcal{B}(0)},
\end{equation}
with the product form preserved (i.e.\ the state is no longer entangled) and therefore
decoherence is suppressed: $|\mathcal{L}(t)| \propto |\langle\mathcal{B}(0) | \hat{T}_u^\dagger(t) \hat{T}_u(t) | \mathcal{B}(0)\rangle|
= |\langle\mathcal{B}(0) | \hat{\mathds{1}} | \mathcal{B}(0)\rangle| = 1$.

\subsection{Interaction and bath Hamiltonians}

We consider the situation where the central spin interacts with a spin-1/2 bath (e.g.\ $^{29}$Si impurities) primarily through the contact hyperfine interaction:\cite{DeSousa2003_1}
\begin{equation}
\hat{H}_{\text{int}} = \sum_a J_a \hat{{\bf S}} \cdot  \hat{{\bf I}}_{a}
\label{Eq:Hint},
\end{equation}
where $\hat{{\bf S}}$ represents the central electron spin, $J_a$ is the strength of the contact hyperfine interaction and $a$ labels the bath spins $\hat{{\bf I}}_{a}$. 
The bath Hamiltonian consists of nuclear Zeeman terms and dipolar coupling among bath spins:
\begin{eqnarray}
\hat{H}_{\text{bath}} &=& \hat{H}_\text{D} + \hat{H}_\text{NZ}, \nonumber\\
\hat{H}_\text{NZ}     &=& \sum_{a} \gamma_N B \hat{I}_a^z, \nonumber\\
\hat{H}_\text{D}      &=& \sum_{a < b} \hat{{\bf I}}_{a} {\bf D}({\bf r}_{ab}) \hat{{\bf I}}_{b},
\label{Eq:Hbath}
\end{eqnarray}
where $\gamma_N$ is the gyromagnetic ratio of bath spins and ${\bf D}({\bf r}_{ab})$ is
the dipolar tensor coupling bath spins $a$ and $b$ separated by ${\bf r}_{ab}$, with components
$D_{i j}({\bf r}) = \mu_0\hbar\gamma_N^2(\delta_{i j}/r^3 - 3 r_{i} r_{j} / r^5 ),$
where $i,j = \{x,y,z\}$, $\delta_{i j}$ is the Kronecker delta and $\mu_0 = 10^{-7}$~NA$^{-2}$
is the magnetic constant divided by $4\pi$.\cite{Schweiger2001}

\subsection{Example central system with OWPs}

The dephasing properties of the central spin qubit and bath are extremely well studied for the case of a
spin-$1/2$ qubit.\cite{DeSousa2003_1,Witzel2005,Witzel2006,Yao2006,Yang2008,Yang2008E,Yang2009}
Good agreement with experiment has been achieved by cluster-based methods
such as the CCE,\cite{Yang2008_2008E_2009}
which decompose the bath dynamics into products of contributions from 
clusters of 2, 3, or more interacting bath spins as illustrated in \Fig{Fig:Entry}(b).
But simple spin-$1/2$ systems do not have OWPs.

For the donor systems, however, the central spin Hamiltonian
$\hat{H}_{\text{CS}} \simeq \omega_0 {\hat S}_z + A\hat{\bf I}_h \cdot \hat{\bf S},$
contains the usual Zeeman term arising from the external magnetic field ($\omega_0 = \gamma_e B$, where
$\gamma_e$ is the electronic gyromagnetic ratio) and also a significant hyperfine coupling ($A$) of the host spin
$\hat{\bf I}_h$ to the electron. For example, for the bismuth donor, $I_h=9/2$ and $A=1.475$~GHz, thus the mixing between
host nuclear and electronic spins becomes substantial for $B\simeq 0-0.3$~T as seen in \Fig{Fig:Entry}(a).

Details of the mixing of states and the corresponding energy levels and transition probabilities were obtained analytically in Refs.~\onlinecite{Mohammady2010,Mohammady2012}.
As illustrated in \Fig{Fig:Entry}(a), there are a total of $2(2I_h+1)$ quantum states (e.g.\ 20 levels
for bismuth with $I_h=9/2$, 8 for arsenic which has $I_h=3/2$). At high magnetic fields, the Zeeman states
$\ket{i} = \ket{m_S,m_{I_h}} $, $ m_S = \pm 1/2, m_{I_h} = -I_h,-I_h+1,\dots,I_h$, provide good quantum numbers. At lower fields, a new adiabatic set of states
$\ket{i} \equiv \ket{\pm ,m} $ must be employed, since $m_S$ and $m_{I_h}$ are not good quantum numbers, but $m=m_S+m_{I_h}$ is. 
The relation between the Zeeman basis and the adiabatic basis is given by:
\begin{eqnarray}
&\ket{+,m} &=  \ \ \cos{\tfrac{\beta_m}{2}} \ket{\tfrac{1}{2}, m-\tfrac{1}{2}} + \sin{\tfrac{\beta_m}{2}} \ket{-\tfrac{1}{2}, m+\tfrac{1}{2}} \nonumber \\
&\ket{-,m } &=  -\sin{\tfrac{\beta_m}{2}} \ket{\tfrac{1}{2},m-\tfrac{1}{2}} + \cos{\tfrac{\beta_m}{2}} \ket{-\tfrac{1}{2}, m+\tfrac{1}{2}}, \nonumber \\
\label{Eq:mixed}
\end{eqnarray}
for all states except the two states with $|m|=I_h+S$ (i.e.\ states 10 and 20 in \Fig{Fig:Entry}(a))
which remain unmixed at all fields. All other states mix with one other, forming doublets of constant $m$.
The transformation between the Zeeman basis and adiabatic basis is given by simple rotation
matrices ${\bf R}^T_y(\beta_m)$ and ${\bf R}_y(\beta_m)$.\cite{Nielsen2010}
Defining parameters $X_m=I_h(I_h+1)-m^2+1/4$ and $Z_m \simeq m+\frac{\omega_0}{A}$,
the angle of rotation is
$\beta_m = \tan^{-1}[X_k/Z_k]$.

For our system of interest, the Zeeman energy of the central system $\omega_0$ dominates
over typical system-bath couplings $J_a$. This motivates a pure dephasing model
(i.e.\ keeping only terms which don't depolarize the states of the central system)
whereby the bath dynamics is governed by effective Hamiltonians depending on the state of the
central system:
$
\hat{h}^{(i)} = P_i \sum_a J_a \hat{I}^z_a + \hat{H}_{\text{bath}}.
$
The key parameter of interest is:
\begin{equation}
P_i(B) \equiv \langle i | {\hat S}_z | i\rangle= \cos{\beta_m},
\label{Eq:theP}
\end{equation}
which is the expectation value of the electron spin $z$-projection; it is no longer fixed at either $m_S=1/2$
or $m_S=-1/2$ as for an unmixed qubit, but is instead a 
strongly field-dependent quantity $P_i(B) \in [-1/2:1/2]$.

OWPs correspond to operating the qubit at particular $B$ values where $P_u(B) \simeq P_l(B)$.
They provide a highly effective method of
mitigating decoherence as illustrated in \Fig{Fig:Entry}(d).
Recently, their importance has been recognised for certain donors in silicon such as arsenic
or bismuth. They were investigated theoretically in
Refs.~\onlinecite{Mohammady2010,Mohammady2012,Balian2012,Balian2014} and 
also experimentally,\cite{Wolfowicz2013,Balian2014} extending the electronic spin coherence time from $0.5$~ms to $100$~ms for an ensemble of donors in natural silicon.
OWPs have been extensively investigated for classical field noise,\cite{Vion2002,Martinis2003,Makhlin2004,Makhlin2004_1,Falci2005,Ithier2005,Mohammady2012,Cywinski2014}
and in Ref.~\onlinecite{Cywinski2014} dynamical decoupling was also studied,
but Refs.~\onlinecite{Balian2012,Balian2014} considered quantum baths consisting of nuclear spins.

The main eight OWPs for the bismuth system, occurring for $B < 0.2$~T are associated with avoided crossings.
The transitions for four of these are shown by the arrows in \Fig{Fig:Entry}(a), while the other four correspond to forbidden
transitions close by. The OWP we consider in this work is for the $\ket{14} \to \ket{7}$ transition which
occurs at magnetic field $B_\text{OWP} = 799$~G. In practical realisations, OWPs
have become closely associated with field values where the derivative of the transition frequency
$f = (E_u - E_l)/2\pi$ with respect to magnetic field vanishes. These are refereed to as
``clock transitions'',\cite{Wolfowicz2013} and correspond to suppression of classical field
fluctuations, such as those arising from instrumental noise.\cite{Mohammady2012}
The OWPs we consider are associated with reduced decoherence from quantum spin baths.
Details of the difference between OWPs and clock transitions
are given in \App{App:dfdB}.

\begin{figure}[t!!!]
\includegraphics[width=3.375in]{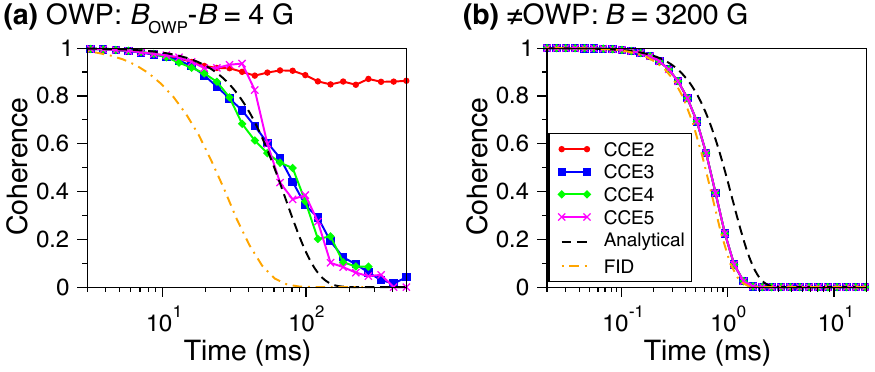}
\caption{(color online) Shows quantum many-body calculations of the Hahn echo ($N=1$) 
using the cluster correlation expansion (CCE) method. {\bf (a)} Near OWPs, calculations using a bath of independent spin pairs only (red, CCE2) 
do not even predict a finite decay time but, surprisingly, calculations with clusters of three spins (blue, CCE3) are already well-converged.
The dashed lines used \Eq{Eq:T2eq}, a closed-form equation derived from the short time behavior, found in Ref.~\onlinecite{Balian2014} to yield good agreement with experiments;
this indicates that three-cluster results too give good agreement with measurements.
Higher order CCE can encounter numerical divergences (which can be attenuated by ensemble averaging);
this accounts for the discrepancies with CCE5.
{\bf (b)} Far from the OWP, independent pairs (CCE2) already give results
in good agreement with CCE3-5 as well as experiments.
The free induction decay (FID) is also shown for comparison.
Note that \Eq{Eq:T2eq} approximates the decay by a pure Gaussian.
CCE calculations were performed for a bismuth donor in
natural silicon for $B$ along $[100]$ and 
the $\ket{14} \to \ket{7}$ transition for which $B_\text{OWP} = 799$~G.
In (a), $B=795$~G while for (b), $B=3200$~G.}
\label{Fig:Convergence}
\end{figure}

\section{Results: low-order dynamical decoupling}

\subsection{Hahn spin echo (CPMG1)}

Understanding of decoherence for such mixed systems nevertheless remains incomplete.
In Refs.~\onlinecite{Balian2012,Morley2013,Balian2014}, CCE2 calculations
were carried out to obtain coherence times for allowed and forbidden
electron spin resonance (ESR) transitions coupling different pairs of states $|u\rangle \to |l\rangle$.
These CCE2 calculations gave excellent agreement with experiment over most regimes. 
However, in the vicinity of the OWPs (where $P_u \simeq P_l$), the CCE2 Hahn echo decay failed to converge and no
decay was obtained other than initially, for
a short time. Single-central spin free induction decay (FID), in contrast, gave finite decays at all magnetic fields.

In Ref.~\onlinecite{Balian2014}, an analytical expression estimating $T_2$ as a function
of $B$ was obtained, by inspection of the short time behavior of the form of the FID 
decays (which can be given analytically for each pair cluster):
\begin{equation}
T_2(B) \simeq \overline{C}(\theta) \frac{\left(|P_u| + |P_l|\right)}{{\left|P_u-P_l\right|}}.
\label{Eq:T2eq}
\end{equation}
The magnetic field dependence is wholly contained in the $\frac{\left(|P_u| + |P_l|\right)}{{\left|P_u-P_l\right|}}$ envelope,
while the prefactor $\overline{C}(\theta)$ depends only on magnetic field orientation, the density of nuclear spins and their gyromagnetic ratio,
but is independent of the strength of $B$.\cite{Balian2014}

This simple closed-form equation gave remarkable and accurate quantitative
agreement with experiment in all regimes, spanning orders 
of magnitude changes in $T_2$, whether in the unmixed limit of a spin-$1/2$, or for certain transitions which are ESR forbidden at high fields, or at OWPs.
The universal validity of \Eq{Eq:T2eq} is worthy of discussion. 
Farther than about $100$~G from the OWP, and where CCE is converged at CCE2,
there is little difference between single-spin FID and Hahn echo decays;
thus, it is not surprising that an equation obtained by considering the independent
spin-pair contribution to FID can accurately model the Hahn echo experiments.
Its validity within the OWP regions, however, is not yet fully understood.
In particular, it remains unclear why a single $\overline{C}(\theta)$ prefactor suffices
to accurately estimate experimental $T_2$, whether very far or very close to OWPs; and to describe different
OWP regions (of which there are 16 for the bismuth system, with $P_{u,l}$ values varying by close to an order of magnitude).

In the present work however, \Eq{Eq:T2eq} is employed simply to simulate the expected experimental
behavior. In \Fig{Fig:Convergence}, we present converged CCE calculations for the Hahn spin echo (CPMG1) near the OWP.
We show that including three-spin clusters (CCE3) give converged results while qubit-bath correlations from only spin pairs
(CCE2) give little decay (red line) except at short timescales.
However, it can also be seen that all orders have similar short time behavior and that the inclusion of the three-clusters
in effect recovers the short time behavior of the pair decays.
Hence, the converged CCE agrees well with \Eq{Eq:T2eq}, 
which was derived from the early time decay of correlations from pairs and hence accounts
for the experimental behavior observed in Ref~.\onlinecite{Wolfowicz2013}.
This is one of the key results of this work. Details of the CCE calculations are given in \App{App:CCE}.

\subsection{CPMG$N$ with few pulses $N \simeq 2-20$}

\begin{figure}[t]
\includegraphics[width=3.375in]{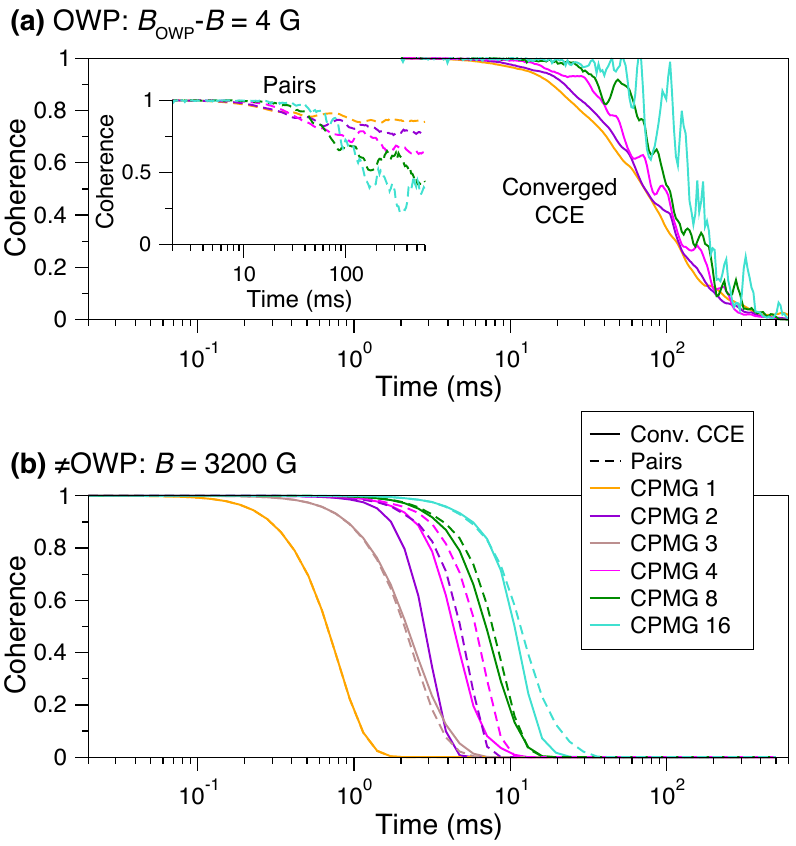}
\caption{(color online) Shows dependence of the coherence on the number of dynamical decoupling pulses $N$,
near {\bf(a)} an optimal working point (OWP) and {\bf (b)} far from an OWP, for modest numbers of $N$.
{\bf(a)} For $B$ close to $B_{\textrm{OWP}}$, the $T_2$ times show comparatively little response to dynamical decoupling. 
Further, even though the initial coherence is extended with increasing $N$, the decays become ever more oscillatory.
For low $N$, the independent pairs contribution is largely eliminated.
{\bf Inset of (a)}: Showing complete suppression of the independent pairs contribution near an OWP;
but showing also its gradual revival as $N$ increases.
{\bf(b)} In contrast, far from the OWP,
substantial (order of magnitude) enhancement of the $T_2$ time by dynamical decoupling is achieved with a moderate (preferably even)
number of pulses. Decays for independent pair contributions (dashed lines, CCE2) and the converged quantum many-body numerics
(solid lines, CCE4) are also compared, indicating that as $N \gtrsim 10$, once again, the independent pair contribution is sufficient.
Parameters are as in \Fig{Fig:Convergence} and for CPMG$N$ sequences.
The converged CCE in (a) corresponds to CCE3.}
\label{Fig:Decays}
\end{figure}

In \Fig{Fig:Decays}, we present comparisons of the response to dynamical decoupling near and away from an OWP 
by means of converged CCE calculations in both cases for $N$ up to 16.
One notable feature of the comparison is the insensitivity of OWP behavior
to low numbers of pulses, in sharp contrast to the $\ne$OWP regime
where there is a factor of 3 ``jump'' in $T_2$ from CPMG1 to CPMG2;
while for larger $N$, we find $T_2 \sim N$ as seen in \Fig{Fig:Entry}(c).
However, OWPs are extremely effective at suppressing decoherence:
for the point shown near the OWP, $T_2 \simeq 100$~ms already at
CPMG1, while away from the OWP, to obtain comparable values, $N\simeq 100$ pulses are required as shown in Section~V.

Previous studies, including a recent study of the ESR dynamics of a phosphorus donor at $X$-band frequencies
(a system without OWPs for electron qubit decoherence),\cite{Ma2014} observed a sharp increase 
in the coherence time between CPMG1 and CPMG2. The spin pair contributions
were also suppressed, requiring many-body correlations for convergence and thus exposing the latter.\cite{Ma2014}
Nevertheless, CCE2 was shown to still give a reasonable approximation to the
magnitude of the observed $T_2$ time, for both CPMG1 and CPMG2.
In the case away from an OWP, the FID is very similar to CPMG1.
This is in contrast to the OWP, where CCE2 gives no decay at all, while the FID gave 
decay curves comparable to converged CCE3 (and \Eq{Eq:T2eq}).
Thus, there is a drastic change from FID to CPMG1 at OWPs; in contrast,
for regimes away from an OWP, there is little change between FID and CPMG1,
but a strong enhancement for CPMG$N$ with $N>1$.

The quantum numerics do evidence a clear dependence of the pair contribution on pulse number $N$. For example, in the inset of \Fig{Fig:Decays}(a),
we have shown that, for a given field $B$ in the vicinity of the OWP,
as $N$ increases to $N\simeq16$, the pair contribution once again gives significant decay.
To suppress decay for $N=16$ one must choose a value of $B$ even closer to the OWP.
In fact this is one of the main findings of the present work: whether at OWPs or far from OWPs,
our comparisons between many-body CCE3-5 and calculations involving only
pairs show that increasing $N$ gradually restores the importance of the pair contribution,
relative to $N=1$ or $N=2$, where many-body effects are seen to make the dominant contribution.
In that case, and whenever independent pairs are dominant, we can employ well established,
two-state analytical pseudospin models of the qubit-bath dynamics.\cite{Liu2007,Zhao2012}

\section{Analysis: pseudospin models}
\label{Sec:pairsmain}

We now proceed to analyse correlations from independent pairs in order to obtain insight on the effect of
dynamical decoupling near and far from OWPs.
As described above, assuming pure dephasing justified by $\omega_0 \gg J_a$,
for the case of a pair of bath spins, the joint system-bath dynamics reduces to a simplified 
two-state form for each of the two qubit states (upper and lower)
and is governed by effective Hamiltonians:
\begin{equation}
 \hat{h}_{u,l}= \frac{1}{4}(\Delta_{u,l} \hat{\sigma}_z + C_{12} \hat{\sigma}_x),
\label{Eq:pseudospin}
\end{equation}
where $\Delta_{u,l}$ is a system-dependent detuning and $C_{12}$ is the strength of the intrabath
secular dipolar interaction.
The Pauli matrices are written in the non-interacting bath basis $\{ \ket{\downarrow \uparrow},\ket{\uparrow \downarrow} \}$,
where $\uparrow$ ($\downarrow$) denotes spin up (down) for each of the two bath spins and we have ignored the
trivial states $\ket{\uparrow \uparrow}$ and $\ket{\downarrow \downarrow}$ which cannot flip-flop.
The detunings in our case are $\Delta_{u,l} = P_{u,l} (J_1 - J_2)$, where $J_1$ 
and $J_2$ are the hyperfine coupling strengths for the two bath spins.
The system-bath evolution is then obtainable analytically from the eigenvalues 
$\omega_{u,l} = \tfrac{1}{4}\sqrt{(\Delta_{u,l} )^2+(C_{12})^2}$ and eigenvectors of
the above effective Hamiltonians.

After preparing the initial qubit superposition, the CPMG$N$ pulse
sequence can be summarized as $(\hat{T}(\tau) - \pi - \hat{T}(\tau))^N$, with final evolution time $t=N\tau/2$.
The unitaries $\hat{T}(\tau)$ represents free evolution and $\pi$ denotes the refocusing pulse which flips
between $\ket{u}$ and $\ket{l}$: $\ket{u}\bra{l} + \ket{l}\bra{u}$ but leaves all other central states unperturbed.
In the so-called pair-correlation approximation,\cite{Yao2006} the coherence decay is simply given by
$\mathcal{L}(t) = \prod_k |\mathcal{L}_k^{(N)}(t)|$, where $|\mathcal{L}_k^{(N)}|$
is the decay contribution from the $k$-th spin pair and the product is over
all spin pairs in the bath.

For CPMG$N$, if 
\begin{equation}
\ket{\mathcal{B}_{u,l}(t)} = \hat{T}^{(N)}_{u,l} \ket{ \mathcal{B}(0) },
\label{eq:time1}
\end{equation}
we can write for the Hahn spin echo case (i.e.\ CPMG1):
\begin{equation}
 \hat{T}^{(1)}_{u,l}= A_0 \hat{\mathds{1}} -i {\bf{A}}_{u,l} \cdot  \hat{\boldsymbol\sigma},
\label{unitary2}
\end{equation}
where ${\bf A}_{u}=(A_x,A_y,A_z) $ and $\hat{\boldsymbol\sigma}$ is the vector of Pauli matrices in the
bath basis: $\{ \ket{\downarrow \uparrow},\ket{\uparrow \downarrow} \}$.
 The ${\bf A}_{u,l} $ components depend on time and can easily
 be given explicitly in terms of the pseudospin parameters; for example:
$A_0(t)=\cos{\omega_u t} \cos {\omega_l t} - \sin{\omega_u t} \sin{\omega_l t} \cos{(\theta_u-\theta_l)}$ ;
and
$A_y(t)=-\sin{\omega_u t} \sin {\omega_l t} \sin{(\theta_u- \theta_l})$.
where  $\theta_{u,l}=\tan^{-1}\left(C_{12}/\Delta_{u,l}\right)$.
In fact, the only term which is not invariant w.r.t. $u \leftrightarrow l$ is $A_y$
and thus ${\bf A}_{l}=(A_x,-A_y,A_z) $.
The coherences
$ |\mathcal{L}^{(N)}(t)| \propto
| \langle \mathcal{B}(0) | \hat{T}^{\dagger(N)}_l \hat{T}^{(N)}_u | \mathcal{B}(0)\rangle|$ are obtained simply from
$\hat{{L}}^{(N)}(t) \equiv \hat{T}^{\dagger(N)}_l \hat{T}^{(N)}_u$.

For both CPMG1 and CPMG2, the unitarity of the evolution of
upper relative to lower states is broken by a term $\propto A_y$. For CPMG1,
\begin{equation}
\hat{L}^{(1)}(t)= \hat{\mathds{1}} -2i A_y \hat{\sigma}_y \hat{T}^{(1)}_u.
\label{CPMG1}
\end{equation}
We can consider higher sequences; since $\hat{T}^{(2)}_u=\hat{T}^{(1)}_u\hat{T}^{(1)}_l$ and $\hat{T}^{(2)}_l=\hat{T}^{(1)}_l\hat{T}^{(1)}_u$, we obtain for CPMG2
\begin{equation}
 \hat{L}^{(2)}(t)= \hat{\mathds{1}} - 4 i A_y (A_z \hat{\sigma}_x-A_x \hat{\sigma}_z)  \hat{T}^{(2)}_u.
\label{CPMG2}
\end{equation}
Both the above general expressions apply equally to either OWP or the $\ne$OWP regimes. 
The only important difference between these regimes is that
$ \theta_u \to \theta_l$ for the approach to an OWP
and $ \theta_u = \pi-\theta_l$ for the spin away from the OWP.
Alternatively, from the explicit expressions for the components of
${\bf A}_{u,l} $, we see that the OWP condition is $A_y \to 0$; since $A_y$ is the prefactor to both 
the above expressions, CPMG1 and CPMG2 are equally suppressed at OWPs.

For the thermal initial bath states $ \ket{\downarrow \uparrow}$ or $ \ket{\downarrow \uparrow}$,
the temporal coherence decays for the $k$-th spin pair of the bath is
$ |\mathcal{L}_k^{(N)}(t)| =| \bra{ \downarrow \uparrow } \hat{L}_k^{(N)}(t) \ket{ \downarrow \uparrow } |
=| \bra{ \uparrow \downarrow } \hat{L}_k^{(N)}(t) \ket{\uparrow \downarrow} |$. The full decay is given by
\begin{equation}
\mathcal{L}(t) = \prod_k |\mathcal{L}_k^{(N)}(t)|.
\label{Eq:CCE2}
\end{equation}
We can easily obtain the coherence decay envelopes for CPMG1 in general, assuming pulse interval $\tau$:
 \begin{equation}
|\mathcal{L}^{(1)}(t=2\tau)|^2= 1- 4 A_{y}^2 A_{0}^2,
\label{Eq:CPMG1EA}
\end{equation}
where we drop the $k$ label for convenience and $A_0\equiv A_0(\tau)$, $A_y\equiv A_y(\tau)$.
For arbitrary even numbers of pulses, CPMG$N$ such that $N/2$ is an integer,
\begin{equation}
\mathcal{L}^{(N)}(t=2 N \tau)= 1- \frac{2 A_{y}^2}{A_{y}^2 + A_{0}^2} \sin^2 \left[\frac{N\phi(\tau)}{2}\right],
\label{Eq:CPMG2N}
\end{equation}
where $\cos \phi(\tau)=A_0(2\tau)$. An equivalent expression was obtained in Ref.~\onlinecite{Zhao2012}.
Both expressions \Eq{Eq:CPMG1EA} and \Eq{Eq:CPMG2N} are equally valid for both regimes (OWP and $\neq$OWP).

\subsection{OWP limit }

The only important difference between these regimes is that
$ \theta_u \to \theta_l$ for the approach to an OWP
and $ \theta_u = \pi-\theta_l$ for the spin away from the OWP.
Alternatively, from the explicit expressions for the components of
${\bf A}_{u,l} $, we see that the OWP condition is $|A_y| \to 0$.
Thus, the suppression of qubit-bath correlations from pairs for OWPs is of the same order
for CPMG1, CPMG2 or any other even-pulsed CPMG: for all bath spin pairs equally, the decay due to correlations from an independent pair
uniformly tends to zero as $(A_{y,k})^2 \to 0$ for the $k$-th pair as $B \to B_{\textrm{OWP}}$.

The dependence on $N$ is entirely contained in the $\sin^2 {N\phi(\tau)/2}$ term. If $N\phi(\tau) \ll 1$
then increasing $N$ has a strong amplifying effect on the signal,
while if $N\phi(\tau) \gg 1$, increasing $N$ simply results in oscillatory behavior.
 Near OWPs, from the expression for $A_0(2\tau)$, we see that if $\theta_u=\theta_l$,
$\phi(\tau)/2 \simeq (\omega_u+\omega_l)\tau$. Hence we only expect a
 response to dynamical decoupling if $\tau$ is sufficiently small (i.e.\ if $\tau \lesssim (\omega_u+\omega_l)^{-1}$).

\subsection{$\ne$OWP limit}

In contrast, for CPMG away from an OWP, the $A_y^2$ prefactor is still there, but is not small.
The origin of the suppression of correlations from independent pairs for small numbers of pulses is more subtle
to analyse with the pseudospin model.
For CPMG2 ($\ne$OWP limit), we obtain:
\begin{equation}
|\mathcal{L}^{(2)}(t)|^2 = 1 - 64 A_{y}^2 A_{0}^2 A_{x}^4.
\label{Eq:CPMG2E}
\end{equation}
The large jump in $T_2$ from CPMG1 to CPMG2 was also analysed in Ref.~\onlinecite{Ma2014}.
In the notation of Ref.~\onlinecite{Ma2014}, we see that 
for CPMG1, the decay envelope is of order $n_x^2$, while for
CPMG2 it is of order $n_x^6 n_z^2$, where 
$n_x= \sin{\theta_u} = \sin{\theta_l}$ while
$n_z= \cos{\theta_u} =- \cos{\theta_l}$. Since the bath spans all
angles $|\theta_{u,l}| =[0,\pi/2]$ one cannot a priori assume  
$\sin{\theta_{u,l}}$ is small.
However, previous numerical 
studies support the idea
that those spin pairs which have $|J_1 -J_2| \gg |C_{12}|$ (i.e.\ are strongly coupled to the central system)
and therefore small pseudospin angles, dominate the Hahn echo
contribution.\cite{Balian2014}
For CPMG2, such strong-coupled spin pairs are  
strongly suppressed, and so $T_2$ becomes dominated by more weakly coupled spin pairs
which are less effective in decohering the qubit.

\section{Results: CPMG$N$ with large $N$}

\begin{figure}
\includegraphics[width=3.375in]{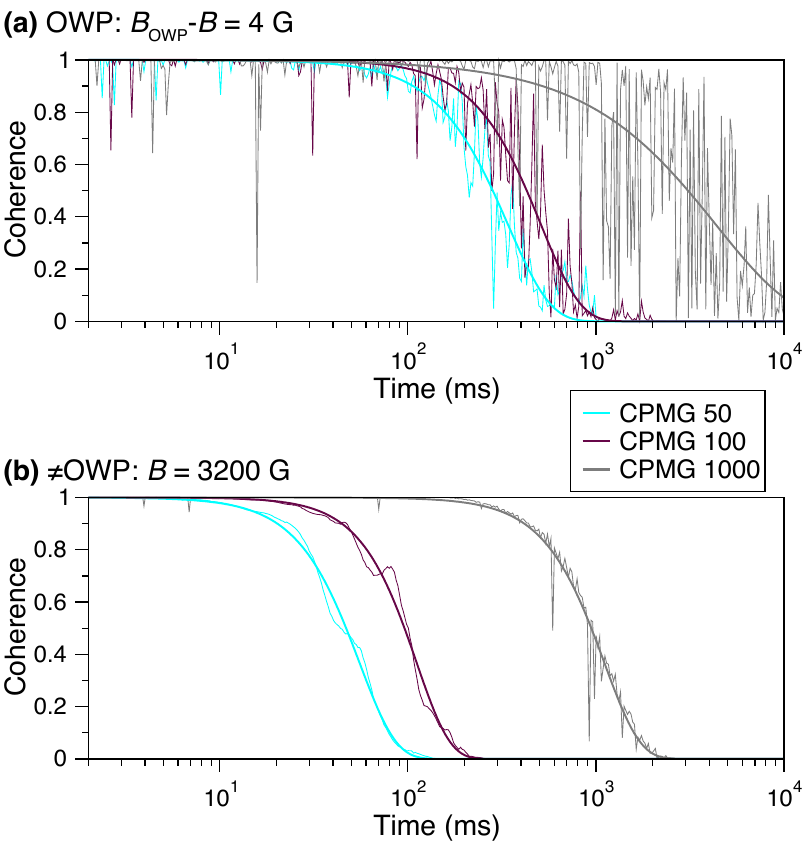}
\caption{(color online) Shows behaviors of coherence decays for large numbers ($N$) of dynamical decoupling pulses 
 {\bf(a)} near and {\bf(b)} far from  OWPs;
as shown in \Fig{Fig:Decays}(a), in this regime, correlations from independent pairs once again dominate the decays
in all regimes so CCE2 is converged and plotted. The behavior at OWPs is now sensitive to $N$ but
the decays here become increasingly oscillatory 
as $N$ and $T_2$ both become large; we attribute this to large numbers of bath spin-pair frequencies becoming resonant with the pulse spacing. It indicates the behavior one might expect in a single-shot single spin study. The smooth
lines are fits to the decays and indicate the behavior expected after ensemble averaging. 
Parameters are as in \Fig{Fig:Convergence}.}
\label{Fig:HighCPMG}
\end{figure}

For large $N$, decays from independent pairs only (CCE2) are restored as well as the sensitivity to dynamical decoupling at OWPs.
Even for $N=16$ (\Fig{Fig:Decays}) we see that the initial period of no decay $\mathcal{L}(t) \sim 1$ is prolonged.
For larger $N$ (\Fig{Fig:HighCPMG}), the enhancement of coherence even at OWPs is clear, but however,
the decays become extremely noisy.
The noise can be attributed to the timescales of individual nearby spin clusters and the time interval between pulses.
For these long coherence times ($\sim 1$s) there are very large numbers of resonances.
The CPMG sequence provides a means of amplifying noise from nearby clusters whenever
 pulse intervals become resonant with the characteristic cluster frequency.
 While this makes the CPMG a valuable technique for spin detection,\cite{Zhao2012_1}
large numbers of such resonances are undesirable if the aim is to protect qubit coherence.
In contrast, far from OWPs, the decays for high $N$ remain relatively smooth.
While the noise at OWPs can be mitigated by ensemble averaging, this is likely introduce a
considerable disadvantage in terms of single-shot operation of a single qubit.

\section{Discussion}

In sum, we have seen that a key difference between OWP and $\neq$OWP behaviors
 arises from the $A_y^2 \propto \sin^2 ({\theta_u-\theta_l})$
 prefactor which globally suppresses all independent pair contributions on the approach to an OWP, 
and accounts for the drastic effect at OWPs, but which is independent of $N$ 
and has little effect far from OWPs.
 However, to analyse decays
 resulting from dynamical decoupling one must consider the remainder of the expression in
\Eq{Eq:CPMG2N}, which reflects the dependence on $N$.
The ineffectiveness of dynamical decoupling near OWPs for small $N$ can also be understood with an
intuitive picture considering the relevant timescales of the system.
For dynamical decoupling to be effective, the time interval between pulses ($t/2N$) must be shorter than to the correlation time of the bath $\tau_c$.
Since typical intra-bath interactions are at most a few kHz, $\tau_c \sim 1$~ms.
Near the OWP, $\omega^u \simeq \omega^l$ and $\theta_u \simeq \theta_l$, so the frequency
of the bath noise spectrum ($\sim \omega_{u,l}$) is appreciably higher than $1/\tau_c$
and thus dynamical decoupling becomes ineffective in extending the coherence time $T_2 \gg \tau_c$.
At short times and for high $N$ however ($t/2N < \tau_C$), dynamical decoupling does protect the central system
as evidenced for CPMG16 in \Fig{Fig:Decays}(a) and higher $N$ in \Fig{Fig:HighCPMG}.
In contrast, dynamical decoupling is far more effective in extending $T_2$ away from the OWP and for relatively
small $N$ (\Fig{Fig:Decays}(b)); although the pseudospin frequencies are comparable, the pseudospin
fields are in opposing directions ($\theta_u \simeq \pi - \theta_l$), thus, the frequency
of noise is much slower and becomes comparable to $1/\tau_c \sim 1/T_2$.

Finally, it is important to note that for direct quantitative comparisons between our
dynamical decoupling calculations and experimental ensemble measurements,
inhomogeneous broadening due to $^{29}$Si nuclei might also have to be factored in
(see \App{App:broadening}).

\section{Conclusions}

Understanding the interplay between optimal working points and dynamical decoupling 
involves understanding of the quantum behavior as a function of 
the two limits $B \to B_{\textrm{OWP}}$ and $N \to \infty$ corresponding to
approaching an OWP and simultaneously increasing the number of dynamical decoupling pulses.
An underlying question of physical interest is when decoherence is the result of the magnetic noise 
from independently flip-flopping pairs of spins and when consideration of the many-body nature of the quantum bath is important.
The answer is of practical importance. For one, if decoherence is due to flip-flopping pairs, there are widely used models (such as the analytical pseudospin
expressions in Section~IV) which can be used to accurately calculate decays. Otherwise, more complex full many-body numerics
become essential to simulate and fully understand experimental behaviors.
The clear answer is that for low order dynamical decoupling, the elimination of correlations from independent pairs is so drastic
at OWPs, that many-body numerics is almost indispensable for full understanding and accuracy.
Even away from OWPs, it was shown that many-body effects make a large contribution for $N \lesssim 10$.
However, once $N \to \infty$ we find little difference between independent-pair and many-body results.

For practical applications, by solving for the many-body qubit-bath dynamics to calculate coherence times,
one can hope to identify the best strategy for enhancing the coherence of donor qubits
whilst still keeping the nuclear spin bath of naturally occurring silicon
for its potential technological use.
By operating near OWPs without dynamical decoupling, the maximum achievable $T_2$ is 0.1~s due to inhomogeneous
broadening from the environmental nuclei.\cite{Wolfowicz2013,Balian2014}
For isotopically purified samples in which the nuclear spin bath
is nearly eliminated, $T_2$ at the OWP was measured to be about 1~s
and is limited by decoherence mechanisms involving donor-donor interactions.\cite{Wolfowicz2013}
Therefore, to bridge this single order of magnitude difference in $T_2$ at OWPs without
resorting to isotopic purification, dynamical decoupling should be applied with at least a few hundred pulses.
The effect of dynamical decoupling in extending coherence times near an OWP
is marginal with a moderate number of pulses (up to $N\sim16$)
in contrast to the usual regimes far from OWPs.
For high donor concentrations, the timescale of donor-donor decoherence is comparable to the
$T_2$ obtained in a nuclear spin bath, hence one might also want investigate suppressing
those mechanisms with dynamical decoupling.

However, combining dynamical decoupling with OWPs is not without its drawbacks. As $T_2, N \to \infty$, potentially many
spins in a silicon bath may become resonant with the dynamical decoupling pulse spacing, resulting in very noisy decays
in single central spin realisations. Although ensemble measurements are unaffected by this noise, this means that for single-qubit
operations, if OWPs can be exploited, their extraordinary potential for coherence suppression may be 
sufficient.

\begin{acknowledgements}
We would like to thank Wen-Long Ma for cross-checking the numerical CCE calculations. We acknowledge Gary Wolfowicz, Fern Watson,
Jacob Lang, and John Morton for useful discussions. 
S.~J.~B. is supported by the Stocklin-Selmoni Studentship via the UCL Impact Programme.
\end{acknowledgements}

\appendix

\section{OWPs vs. clock transitions}
\label{App:dfdB}

The OWPs we consider correspond to suppression of decoherence in quantum spin environments.
Similar points of interest are clock transitions or $df/dB=0$ points,\cite{Wolfowicz2013,Mohammady2012} where decoherence arising from
classical field noise is suppressed.
For donor spin systems, the OWP is close to but not exactly at the $df/dB=0$
point, and not all $df/dB=0$ points are OWPs as shown in Ref.~\onlinecite{Balian2014}.
This is because the donor is coupled to the nuclear spin bath primarily
via the electron, and the couplings between the host donor nucleus and bath nuclei
are negligible compared to the electron-bath hyperfine interaction.
It is easy to show that the fields at which $df/dB=0$ satisfy
\begin{equation}
0 =  P_u - P_l + \frac{\delta \Delta m }{1+\delta},
\end{equation}
where $\delta \simeq 10^{-4}$ is the ratio of the electronic to host nuclear gyromagnetic ratios
and $\Delta m$ is the difference in quantum number $m$ between the upper and lower levels.
Since $\delta \ll 1$, the OWP (where $P_u = P_l$) and $df/dB=0$ points {\em nearly} coincide.

It is important to note that all of the transitions with OWPs couple {\em two neighbouring} avoided crossings.
Selection rules were detailed in Ref.~\onlinecite{Mohammady2012}, but all such transitions have $\Delta m =  \pm 1$
which implies that $\langle u | {\hat S}_z| l\rangle=0$. Thus, magnetic field fluctuations
represent pure dephasing noise.
One might also  consider the possibility of creating a superposition of two states 
$|u\rangle$ and $|l\rangle$ at a single avoided crossing; for example, the superposition
$|11\rangle + |9\rangle$ in \Fig{Fig:Entry}(a), at the avoided crossing between these states at $B=0.21$T.
Although the $\ket{11}\to \ket{9}$ transition is never allowed, such a superposition
might be created by a two pulse excitation from level $\ket{10}$.
Both states are at zero energy gradient ($dE_{u,l}/dB=0$) so coherences are to first order insensitive
to dephasing noise; however, as shown in Ref.~\onlinecite{Mohammady2012},
in that case $\langle u | {\hat S}_z| l\rangle\neq 0$ so magnetic fluctuations couple
the states in the superposition and thus coherence is vulnerable to depolarisation by magnetic noise.

\section{Cluster correlation expansion simulations}
\label{App:CCE}

The CCE is a well-established method for accurately calculating the coherence decay $\mathcal{L}(t)$
of a central spin system in a quantum spin bath.\cite{Yang2008_2008E_2009}
In the CCE formulation, the spin bath is decomposed into groups of spins or ``clusters'' and
the closed evolutions of clusters interacting with the central system are combined
to approximate the exact coherence involving the entire spin bath.

The expansion is given by
\begin{align}
\mathcal{L}(t) &= \prod_{ \mathcal{K} } \tilde{\mathcal{L}}_\mathcal{K}(t), \nonumber \\
\tilde{\mathcal{L}}_\mathcal{K}(t) &= \frac{\mathcal{L}_\mathcal{K}(t)}{\prod_{\mathcal{R} \subset \mathcal{K} } \tilde{\mathcal{L}}_\mathcal{R}(t)},
\end{align}
where $\mathcal{K}$ is a subset of the bath and the first product is over all subsets of the bath.
The irreducible or ``true'' correlation term for cluster $\mathcal{K}$, $\tilde{\mathcal{L}}_\mathcal{K}(t)$, is recursively defined as follows.
First, the coherence is calculated by exactly solving for the combined qubit-bath dynamics governed by
the total Hamiltonian in \Eq{Eq:Htotal}, but including only those bath spins contained in $\mathcal{K}$.
Second, the resulting $\mathcal{L}_\mathcal{K}(t)$ is divided by all correlation terms formed out of the proper subsets $\mathcal{R}$ of $\mathcal{K}$.
The usefulness of the CCE becomes clear when truncating the expansion,
\begin{equation}
\mathcal{L}_{[k]}(t) = \prod_{|\mathcal{K}| \leq k} \tilde{\mathcal{L}}_\mathcal{K}(t),
\end{equation}
in which only clusters containing a maximum of $k$ bath spins are included.
The lowest non-trivial order for nuclear spin diffusion is the pairwise correlation approximation ($k=2$ or CCE2),
given by \Eq{Eq:CCE2}, and we calculate up to $\mathcal{L}_{[k=5]}(t)$.
The coherence decay is often considered converged when
$| \mathcal{L}_{[k']}(t) - \mathcal{L}_{[k'+1]}(t) | \ll 1, \forall t$
and for our case this condition is satisfied for up to $k'=3$.

Note that in the CCE method, the pure dephasing approximation is not required, and the interaction
Hamiltonian in general includes terms which depolarize the states of the central system.
Our CCE calculations include the $\hat{S}^{-}\hat{I}^+ + \hat{S}^{+} \hat{I}^{-}$
terms in the hyperfine interaction Hamiltonian \Eq{Eq:Hint}, but we find that these
give small corrections to the case when only $\hat{S}^z \hat{I}^z$ terms are included.
This was expected due to the large mismatch between electronic and nuclear gyromagnetic ratios.

For our simulations, crystal sites of a cubic silicon superlattice
were uniformly populated with $^{29}$Si nuclei ($I=1/2$) with the natural fractional
abundance of 0.0467 and with equal probability of spin-up and spin-down for the initial states.
To calculate the electron-bath hyperfine couplings, we use the Kohn-Luttinger
electronic wavefunction for the bismuth donor in silicon
with an ionization energy of 0.069~eV.\cite{DeSousa2003_1}
The total size of the spin bath is dictated
by the spatial extent of the wavefunction which decays exponentially
with distance from the donor site, and a superlattice of side length 160~\AA (with $10^{4}$ impurities)
gave convergent coherence decays.
Due to cubic decay of the dipolar interaction as the
distance between a pair of $^{29}$Si spins is increased,
it is not necessary to include all spin clusters in the
calculation. At the lowest non-trivial CCE order (CCE2),
spins separated by at most the 4-th nearest neighbor distance in silicon ($\sqrt{11} a_0 / 4$, where $a_0 = 5.43$~\AA)
gave convergent decays away from OWPs.
To choose 3-clusters (i.e.\ including three bath spins), we loop over all sites in
the crystal and add to each 2-cluster only those spins that are
at most separated by $\sqrt{11} a_0 / 4$ from any of the two
spins in the 2-cluster. The same procedure was applied to choose higher-order
clusters, by adding spins to clusters one order down.
In \Fig{Fig:Convergence} and \Fig{Fig:Decays}, the numbers of 2, 3, 4, and 5-clusters were each of order $\approx 10^4$.

\begin{figure}[h!!]
\includegraphics[width=3.375in]{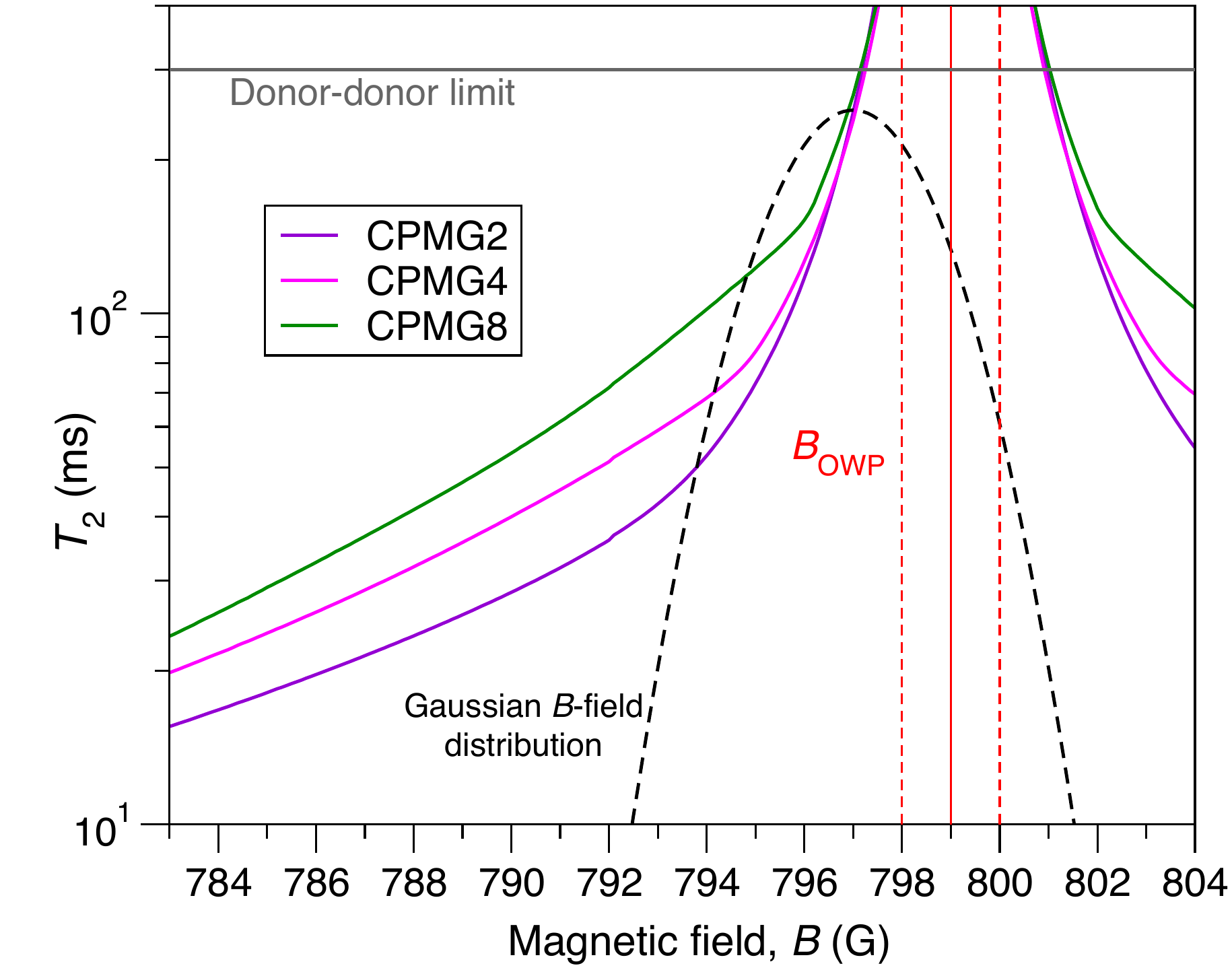}
\caption{(color online) Sharp $B$-field dependence of $T_2$ for various
CPMG orders near an OWP. Inhomogeneous broadening from \tsc{29}Si nuclei
can be incorporated by convolving the decays with a Gaussian $B$-field distribution centred about
$B$ (here centred about $797$~G) and with standard deviation $w \simeq 2$~G (dashed line).
For a donor concentration of $3\times10^{15}$~cm\tsc{-3},
$T_2$ is limited by donor-donor processes at about $300$~ms.\cite{Wolfowicz2013}
The $T_2$ lines were calculated for bismuth donors using the CCE up to 3rd order
and for $B \parallel [\bar{1} 1 0]$. The OWP under investigation is shown in red
at $799$~G.
}
\label{Fig:T2Estimate}
\end{figure}

\section{Effect of inhomogeneous broadening}
\label{App:broadening}

For direct quantitative comparisons between our dynamical decoupling calculations and
experimental ensemble measurements, inhomogeneous broadening might also have to be considered.
As shown in \Fig{Fig:T2Estimate} for various orders of CPMG,
$T_2$ varies sharply with magnetic field over a few G near an OWP.
Inhomogeneous broadening of $B$ due to \tsc{29}Si impurities
has a FWHM of about $4$~G in natural silicon and may
therefore need to be included in the calculation
in order to predict the shape and rate of experimental decays near OWPs.
The broadening can be simulated by convolving the decays $\mathcal{L}_B(t)$
with a Gaussian magnetic field distribution with standard deviation $w \simeq 2$~G:
\begin{equation}
D_B(t) = \frac{1}{w \sqrt{2\pi}}
\int{ e^{\frac{-\left(B-B'\right)^2}{2w^2}} \mathcal{L}_B(t) dB'}.
\end{equation}
Depending on the donor concentration, donor-donor processes
may also need to be included. For example, for a donor concentration
of $3\times10^{15}$~cm\tsc{-3}, $T_2$ near an OWP is limited by direct
flip-flops of the central donor
with other donors in the ensemble.\cite{Wolfowicz2013}
The measured $T_2$ in an isotopically purified sample ranges from $0.2 - 2$~s
($T_2$ for a donor concentration of $3\times10^{15}$~cm\tsc{-3} is 300~ms).
Therefore, care should be taken to include donor-donor processes
very near the OWP (within about 1~G), where nuclear spin diffusion coherence times
are comparable to those of donor-donor processes.

\bibliography{refs}

\end{document}